\documentclass[fleqn,usenatbib]{mnras}

\usepackage{newtxtext,newtxmath}

\usepackage[T1]{fontenc}

\DeclareRobustCommand{\VAN}[3]{#2}
\let\VANthebibliography\thebibliography
\def\thebibliography{\DeclareRobustCommand{\VAN}[3]{##3}\VANthebibliography}

\usepackage{graphicx}	
\usepackage{amsmath}	

\title[ZooBot:3D -- Automated Segmentation]{Deep Learning Segmentation of Spiral Arms and Bars for 600,000 Galaxies in DESI}

\author[A. Spindler et al.]{
Ashley Spindler,$^{1}$\thanks{E-mail: a.spindler@herts.ac.uk}
Mike Walmsley,$^{2,3}$
Karen L. Masters,$^{4}$
Tobias G\'eron,$^{2}$
Izzy L. Garland,$^{5,6}$
\newauthor
B. D. Simmons,$^{6}$
Jürgen J. Popp$^{7}$
\\
$^{1}$ Centre for Astrophysics Research, Department of Physics, Astronomy and Mathematics, University of Hertfordshire, Hatfield, Hertfordshire, AL10 9AB, UK\\
$^{2}$ Dunlap Institute for Astronomy and Astrophysics, University of Toronto, Toronto, ON M5S 3H4, Canada\\
$^{3}$ Jodrell Bank Centre for Astrophysics, Department of Physics \& Astronomy, University of Manchester, Manchester M13 9PL, UK\\
$^{4}$ Department of Physics and Astronomy, Haverford College, Lancaster Avenue, Ardmore, PA 19041 USA\\
$^{5}$ Department of Theoretical Physics and Astrophysics, Faculty of Science, Masaryk University, Kotlářská 2, Brno, 611 37, Czech Republic\\
$^{6}$ Department of Physics, Lancaster University, Lancaster LA1 4YB, UK\\
$^{7}$ School of Physical Sciences, The Open University, Milton Keynes, MK7 6AA, UK
}

\date{Accepted XXX. Received YYY; in original form ZZZ}

\pubyear{2015}

\begin{document}
\label{firstpage}
\pagerange{\pageref{firstpage}--\pageref{lastpage}}
\maketitle

\begin{abstract}
We present a catalogue of segmentation maps identifying the extent of spiral arms and bars of 639,636 galaxies in the DESI Legacy Survey. To produce these maps, we have trained a deep U-net-style neural network using the pixel masks from the Galaxy Zoo: 3D citizen science project. The resulting data products are "soft" segmentation maps, which show the confidence of the model that a pixel lies within the spiral arms or bars of a galaxy. In this paper we detail the sample selection from DESI-LS using the machine classifications from Galaxy Zoo: DESI, the architecture of the U-net model--dubbed ZooBot:3D. We demonstrate the ability of the model to identify spiral arms and bars in a wide range of face-on disks, and identify an emergent ability to identify rings--despite only a small number of ring-type galaxies being present in the training data. Finally, we discuss the practical application of these data products to photometric imaging and IFU spectroscopy. The ZooBot:3D dataset is available for use publicly, and contains the full catalogue presented in this paper, along with cross-matched subsamples for the MaNGA and SAMI IFU surveys.
\end{abstract}




\section{Introduction}
\begin{figure*}
    \centering
    \includegraphics[width=0.75\textwidth]{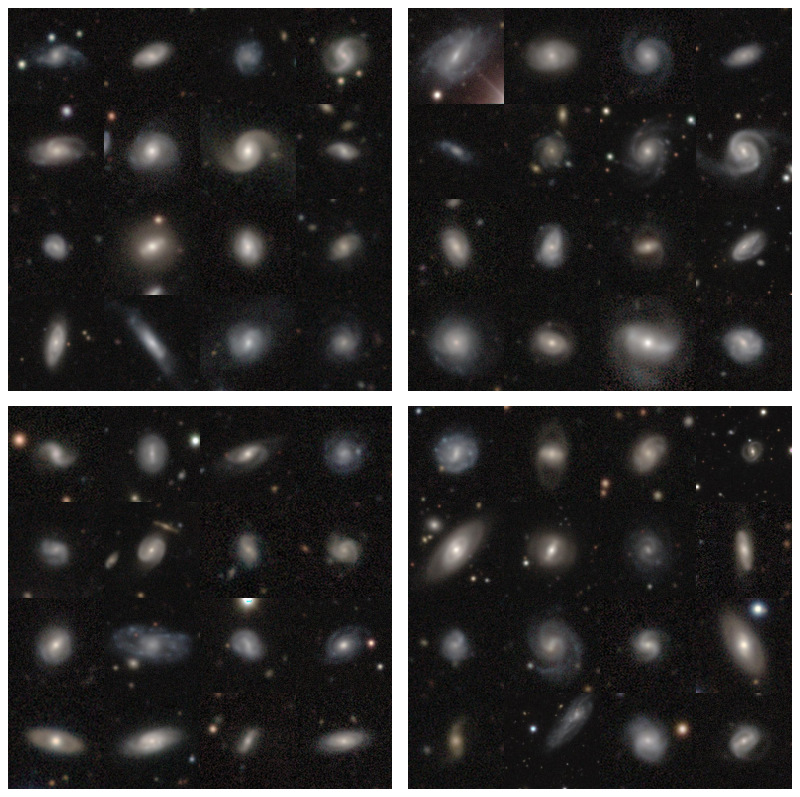}
    \caption{Examples of the DESI images used as inputs for the segmentation model. The galaxies are selected at random from the subsample of GZ: DESI objects detailed in Section \ref{sec:SampSelect}. Each image has an arbitrary pixel scale, having been resampled from native telescope resolution to a $424 \times 424$ pixels with a field of view as described in Appendix A of \citealt{2023MNRAS.526.4768W}. The images have subsequently been cropped with a zoom factor of $0.65$ and resized to $224 \times 224$ pixels.}
    \label{fig:rgb_images}
\end{figure*}

There are many mysteries to solve regarding the evolution and morphology of disk-type galaxies. Astronomers still struggle to pin down the processes that lead to the formation of spiral arms, whether they result from modes in density waves or swing amplification \citep{Sellwood_2022}. Likewise, our understanding of galactic bars is also limited, and things become even more complex when we consider the co-evolution of these structures. Spiral galaxies make up the majority of non-dwarf galaxies in the local universe \citep{2009MNRAS.393.1324B}, and the majority of spirals host galactic bars (e.g. \citealt{Eskridge2000, 2025ApJ...987...74G}). Roughly two-thirds of spirals host two spiral arms and typically have some degree of rotational symmetry.

Spiral structure can be "flocculant" or "grand design" in nature, and can possess a range of pitch angles and pattern speeds \citep{1964ApJ...140..646L, 1982MNRAS.201.1021E, 2009PhT....62e..56B, Sellwood_2022, 2025RNAAS...9..236P}. Spiral arms are known to drive secular evolution processes, thus altering the stellar and chemical make-up, and dynamics of the host disk \citep{1972MNRAS.157....1L, 1972ApL....12...49R, 2002MNRAS.330...35A, 2004ARA&A..42..603K, 2014RvMP...86....1S, 2022A&A...661A..98Y}. Models of density wave patterns show that spiral arms can cause shocks in the cold gas as the arm moves, resulting in gravitational collapse and subsequent star formation \citep{1969ApJ...158..123R}. Stronger spiral arms induce larger shocks, leading to larger concentrations of star formation in the spiral arms \citep{2015MNRAS.446.4155K, 2021ApJ...917...88Y}. The presence of spirals churns the stars in the disks, resulting in flattened rotation curves and smoother density profiles \citep{2015ApJ...799..213B}, while also reducing the metallicity gradients through radial diffusion \citep{2009MNRAS.396..203S, 2015ApJ...808..132H, 2025ApJ...983...57W}.

Likewise, bars also act as significant drivers in the secular evolution of galaxies. At low redshifts, strong or weak bars occupy around $43\%-52\%$ of disk galaxies \citep{2007ApJ...659.1176M, 2008ApJ...675.1194B, 2009A&A...495..491A, 2019MNRAS.488.2175B, 2021MNRAS.507.4389G}, although in some infrared studies this proportion is as high as $80\%$ \citep{Eskridge2000}. The primary mechanism of this evolution is through the transfer of angular momentum from the inner disk to the outer disk and dark matter halo \citep{1972MNRAS.157....1L, 1993RPPh...56..173S, 2003MNRAS.341.1179A, 2013MNRAS.429.1949A}. This momentum transfer funnels gas from the outskirts of the galaxy into the central regions, triggering starbursts and driving the activity of AGN \citep{2013MNRAS.429.1949A, 2013ApJ...779..162C, 2015MNRAS.454.3641F, 2015MNRAS.448.3442G, 2024MNRAS.532.2320G}. Bar activity has also been linked with the shut-down of star formation in disk galaxies, likely due to the higher rate of consumption of cold gas supplies \citep{2020MNRAS.499.1116F, 2021MNRAS.507.4389G, 2011MNRAS.411.2026M}.

In taking steps towards solving the mysteries of spiral galaxies, astronomers have turned to spatially resolved studies of galaxies—namely, large-scale Integral Field Spectroscopy (IFS) surveys like MaNGA (Mapping Nearby Galaxies at Apache Point Observatory, \citealt{2015ApJ...798....7B}) SAMI (Sydney-AAO Multi-object Integral-Field Spectrograph, \citealt{2012MNRAS.421..872C}), CALIFA (Calar Alto Legacy Integral Field Area survey, \citealt{2012A&A...538A...8S}), and AMUSING (All-weather MUse Supernova Integral-field Nearby Galaxies, \citealt{2016MNRAS.455.4087G}). IFS observations allow for detailed studies of a galaxy's spectral and spatial properties, such as how the kinematics, chemical abundances, stellar mass, and gas emission are distributed throughout the galaxy's structure. 

To facilitate decomposed studies of spiral galaxies, \cite{2021MNRAS.507.3923M} ran a citizen science project named Galaxy Zoo: 3D (hereafter GZ:3D) to collect annotations of spiral arms and bars for galaxies in the MaNGA parent sample \citep{Wake_2017} from volunteers. The volunteers were shown three-band images from the Sloan Digital Sky Survey (SDSS, \citealt{1998AJ....116.3040G, 2000AJ....120.1579Y, 2002AJ....124.1810S}), and asked to use annotation tools to mark the extent of the galactic bar and the locations of the spiral arms (in practice, roughly marking the maxima of the spiral mass distribution)\footnote{GZ:3D volunteers also provided annotations for locating galaxy centres and foreground stars, which are not used in this paper.}.

A variety of tools for decomposing galaxy images into their internal structures have been developed, with traditional methods relying on fitting the light profiles to known distributions, such as ellipses for stellar bars. GALFIT \citep{2010AJ....139.2097P} remains a popular approach for decomposing broadband images, while a number of tools for IFS data have also been released in recent years, such as  Bulge-Disc Decomposition of IFU data (BUDDI, \citealt{2017MNRAS.465.2317J}). Many of these approaches and tools are limited by scale, however, as they can be computationally intensive or require manual tuning of individual objects in order to achieve well fitted decompositions.

The rise of machine learning in astronomy has led to some experimentation in using generative models to produce segmentation maps of galaxy structures (see \cite{xu2024surveyingimagesegmentationapproaches} for an overview of segmentation methods in astronomy more generally). In \citet[][hereafter, WS23]{2023arXiv231202908W}, we presented a generative network trained on the GZ:3D spiral arm and bar masks. Our model used a regression approach to predict the fraction of volunteers that included each pixel in the corresponding masks across the plane of the galaxy, and was able to produce detailed, accurate maps of spiral arms and bars. In WS23 surveyed expert astronomers on a comparison of the GZ:3D pixel masks, our machine-generated maps, and segmentation maps from the \texttt{SpArcFiRe} \citep{2014ApJ...790...87D} package--our segmentation maps were preferred over the volunteer masks in $79\%$ of evaluations and over the SpArcFiRe maps in $99\%$ of evaluations.

Other authors have approached the problem of using neural networks to produce galaxy segmentation maps. \cite{2024MNRAS.530.1171C} trained a U-net model on the GZ:3D pixel masks with SDSS and HSC images as the model input. They found that, similar to WS23, the resulting bar masks can effectively recover measurements of bar length. The model used in \cite{2024MNRAS.530.1171C} is comparatively smaller than the model used in WS23--and by extenion, this work--and utilises smaller input images, thus limiting the fidelity of the segmentation products. \cite{2023BAAA...64..253R} also demonstrated a U-net approach to segmentation using the GZ:3D data products, however, their model was trained using a binary classification target, which results in simpler outputs. Neither of these works utilise imaging from DESI.

This work provides a fuller description of the model produced in WS23, and describes a data release composed of spiral arm and bar masks predicted for a large sample of spiral galaxies from Dark Energy Spectroscopic Instrument Legacy Survey (DESI-LS, \citealt{2019AJ....157..168D}). In Section \ref{sec:Data} we describe the data used throughout the work, including the volunteer spiral and bar masks from GZ:3D and imaging from DESI-LS, we also detail the sameple selections made using morphological classifications from Galaxy Zoo: DESI (GZ:DESI, \citealt{2023MNRAS.526.4768W}). Section \ref{sec:ZooBot} expands on the model description from WS23, detailing the architecture, training and prediction procedures. Section \ref{sec:Results} explores the properties of the machine generated segmentation maps, and Section \ref{sec:Conclusions} discusses the conclusions of this work.

\begin{table}
    \centering
    \begin{tabular}{cc}
        Selection Cut & No. of Galaxies\\
        \hline\hline
        Galaxy Zoo: DESI (Parent Sample) & $8,689,370$\\
        Estimated Petrosian Radius $> 2''$ & $4,308,335$\\
        Featured or Disk Fraction $> 0.27$& $1,684,618$\\
        Disk Edge On Fraction $< 0.68$ & $639,636$\\
        \hline
        Has Spiral Arms Fraction $> 0.4$ & $512,216$\\
        Bar Strong + Bar Weak Fraction $> 0.5$ & $230,511$\\
    \end{tabular}
    \caption{This table shows the effect of various selection cuts applied to the GZ-DESI parent sample used in this work. The first column details the cut to be applied, and the second column shows the number of galaxies remaining in the sample after each successive cut. The fractions refer to the machine-generated probabilities for the questions in the Galaxy Zoo decision tree. The counts shown for galaxies with spiral arms or bars are independent of each other, boths cuts are applied to the "Disk Edge On" selection.}
    \label{tab:cuts_table}
\end{table}

\begin{figure*}
    \centering
    \includegraphics[width=1.0\textwidth]{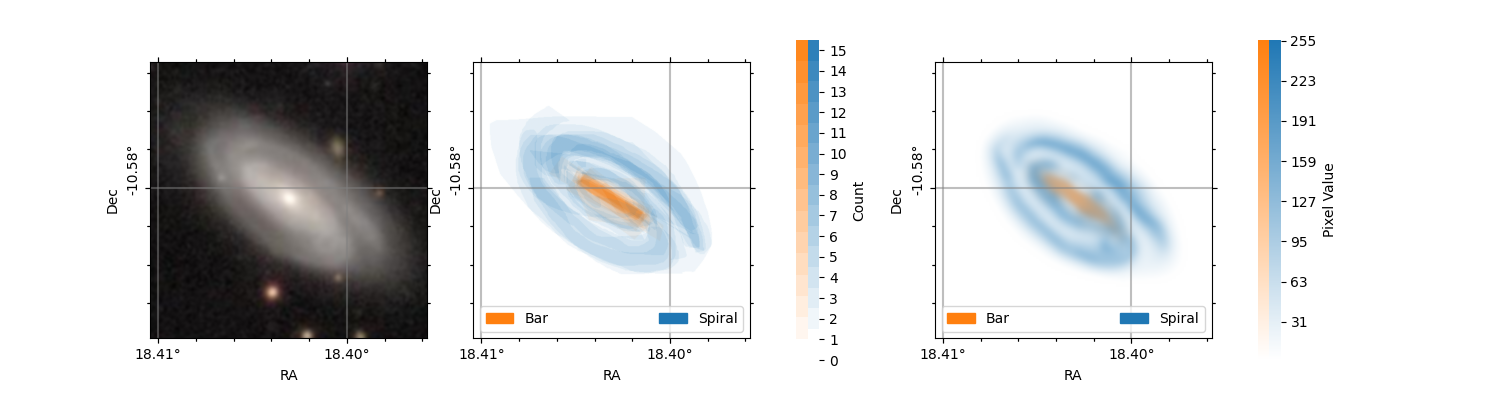}
    \includegraphics[width=1.0\textwidth]{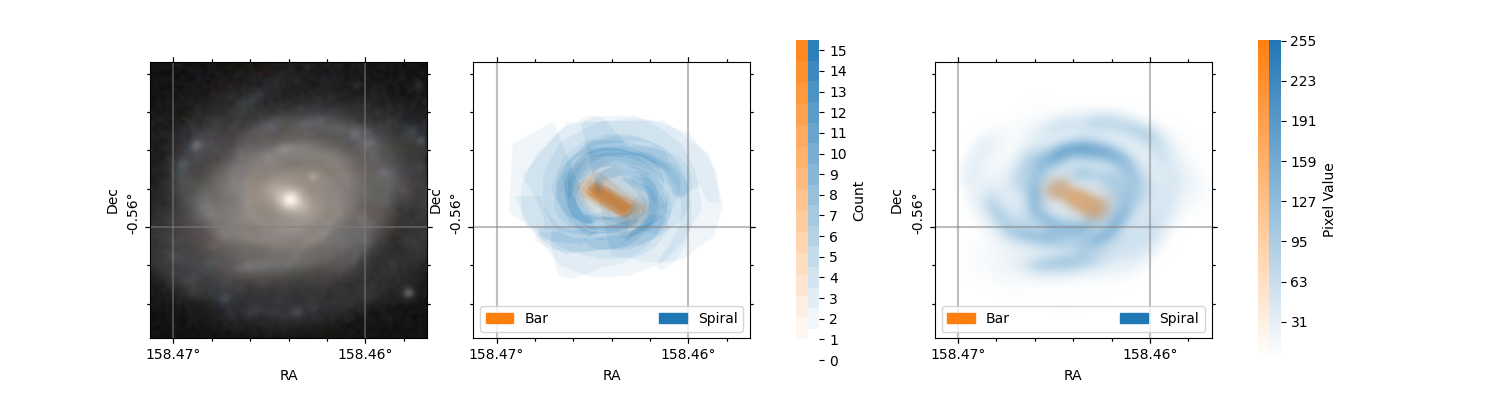}
    \includegraphics[width=1.0\textwidth]{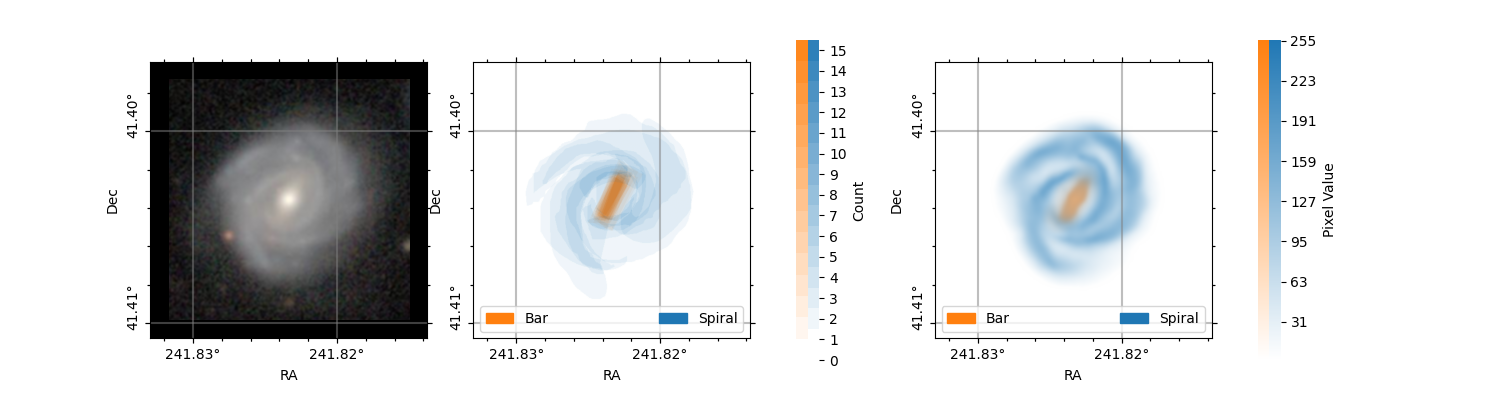}
    \caption{Comparisons of GZ:3D volunteer pixel masks and the output from the segmentation model for three galaxies (MaNGa IDs: $1-103484$, $1-248420$, $1-2511$. DESI DR8 IDs: $270262\_648$, $550292\_354$, $328121\_812$). In the first column we show the DESI image of the galaxy. In the second column we show the GZ:3D volunteer classifications, with the spiral arm masks in purple and the bar masks in red--the pixel intensity denotes the number of volunteers who enclosed the pixel in their mask. The third column displays the output of the segmentation model, with the same colour scheme--here the pixel intensity denotes the confidence of the model that a pixel belongs in the segmentation map.}
    \label{fig:ZB3D_GZ3D_comp}
\end{figure*}

\section{Data}
\label{sec:Data}

\subsection{Galaxy Zoo: 3D}

The training data, in the form of pixel-wise segmentation maps, used in this work are drawn from GZ:3D \citep{2021MNRAS.507.3923M}. In GZ:3D, volunteers were asked to use a polygon drawing tool to label the bars and spiral arms of $7,418$ disk-type galaxies, drawn from the parent sample of MaNGA containing $29,831$ galaxies \citep{Wake_2017}. Each galaxy was labelled by $15$ volunteers, who were presented with the SDSS-I/II $gri$ \citep{1998AJ....116.3040G, 2000AJ....120.1579Y, 2002AJ....124.1810S} image of the galaxy. We recalculate the segmentation maps using the individual volunteer polygon vertices, including self-intersecting areas, to find the number of volunteers who enclosed each pixel in the image. We encode the spiral arm and bar maps as JPEG images using a linear relationship between pixel value and volunteer fraction. A pixel value of $255$ corresponds to all $15$ volunteers enclosing a pixel, and a value of $0$ means no volunteers enclosed a pixel. At training the maps are loaded in as grayscale images and combined into a two-channel object, and the pixel values are scaled to be between $0$ and $1$.

To construct the training sample, we cross-match the $7,418$ galaxies with spiral arm masks from GZ:3D with DESI-LS, finding $6,382$ galaxies in both samples. These galaxies make up the data used to train our model, which we split into separate training/validation/testing sets with a $70\%/10\%/20\%$ split. Note that we require each galaxy in the training sample to have a spiral arm mask from GZ:3D, which means that some barred galaxies with no visible spiral arms are not included in the training sample.

\subsection{Sample Selection}
\label{sec:SampSelect}

Many of the objects in GZ:DESI are not resolved enough to identify internal structures, and so it would not be useful to run the deep learning segmentation model on the full catalogue of some $8.7$ million galaxies. Of those that are well resolved, many will be early-type galaxies or edge-on disks in which spiral arm and bar structures either do not exist or are obscured from view. The first step in producing a catalogue of segmentation maps for DESI sources is selecting which sources the model should be applied to. The following selection cuts are made to be fairly permissive, and we encourage the user to apply more aggressive cuts to achieve particular science goals.

The cuts made to the parent sample from GZ:DESI are summarised in Table \ref{tab:cuts_table}. We first select galaxies with an estimated $50\%$ Petrosian radius of $2''$ in order to limit our sample to galaxies that have resolvable structures. We note that when combined with our cuts based on the GZ:DESI vote fractions below, the vast majority of sources have angular sizes above our cut. Of the $670,000$ face on disks, only $~40,000$ have an angular size smaller than $2''$. The main reason for applying this cut is due to the nature of the training data used, which are drawn from the MaNGA survey—the vast majority of which are very well-resolved, nearby objects.

We then use the automated vote fractions from GZ:DESI \citep{2023MNRAS.526.4768W} to select a "dirty" sample of disk-type galaxies. The reason for allowing for a wider sample is that there may be some objects that might not be fit for dedicated science on spiral arms or bars, but nonetheless produce useful and interesting segmentation maps (e.g. ring galaxies). To select for this sample of disk galaxies, we follow a similar procedure used by other authors when working with the GZ:DESI classifications (see \citealt{2024MNRAS.532.2320G} for example, and \citealt{2021MNRAS.507.4389G} using the earlier Galaxy Zoo DECaLS classifications). We apply cuts based on the predicted vote fraction for \texttt{featured-or-disk} in the \texttt{smooth-or-featured} question and the predicted vote fraction for the \texttt{disk-edge-on} question. We require that the \texttt{featured-or-disk} fraction be greater than $0.27$ and for the \texttt{disk-edge-on fraction to be $< 0.68$}. These cuts, combined with the angular size cut, give us a sample of 639,636 galaxies for which we can predict segmentation maps. We make no cuts based on the vote fractions of the merging question in GZ:DESI, but we note that the \cite{2024MNRAS.532.2320G} selection criteria for merging and disturbed galaxies finds $122,064$ galaxies in our sample.

Further to these cuts, we note that $512,216$ of these galaxies have a \texttt{has-spiral-arms} fraction $> 0.4$, which provides a relatively clean sample of spiral galaxies. A clean sample of barred galaxies can be selected by applying a cut using the sum of the predicted vote fractions for strong and weak bars $> 0.5$, resulting in $230,511$ barred galaxies.

Finally, we construct two catalogues by cross-matching the full MaNGA and SAMI target lists with DESI-LS, to allow us to package segmentation maps for just the galaxies in those surveys. Without any cuts based on the GZ:DESI vote fractions, we find $29,006$ galaxies from the MaNGA parent sample (which includes objects not observed in the final data release) and $5,019$ galaxies from SAMI in the DESI-LS footprint. We predict spiral arm and bar masks for all of these galaxies for completeness; however, users of these data products will need to apply sensible cuts to ensure data quality. It should be noted that spiral arm and bar masks already exist for the MaNGA sample, in the form of the GZ:3D volunteer labelled masks, along with foreground star masks and locations of the galaxy centres. These data products are available as a Value-Added Catalogue to Data Release 17 of SDSS-IV \citep{2022ApJS..259...35A}\footnote{\url{https://www.sdss4.org/dr17/data_access/value-added-catalogs}}.

\subsection{Images}
\label{sec:Images}

In GZ:3D the volunteers were shown the SDSS images of the galaxies they needed to label, and these images would be the natural first choice for training a deep segmentation model. However, imaging from the DESI Legacy Survey are comparable to the SDSS images while revealing more detail in the spiral structures due to greater imaging depth ($r=23.9$ mag for DESI-LS, compared to $r=22.7$ mag for SDSS). Using DESI-LS images in the training phase also allows us to make predictions of segmentation maps across the entire DESI-LS footprint, widening the use case of these data products to large photometric studies, beyond the comparatively small-scale IFS surveys.

Reduced flux measurements of the source galaxies are downloaded from the DESI-LS cutout service in the manner following GZ:DESI. We use the RA and DEC listed in the DESI catalogues for the cutout centres for the main data release of the paper. For the MaNGA and SAMI galaxies (including the MaNGA galaxies used in the training sample), we use the respective RA and DEC from those surveys. The images are downloaded at native telescope resolution ($0.262$ arcsec per pixel), with similar field-of-view as GZ:DESI (see Appendix A in \citealt{2023MNRAS.526.4768W} for further details). The images are then resampled to $424\times424$ pixels, with an arbitrary pixel scale to ensure each galaxy has a similar extent in the image.

During the training of the model, we randomly apply a range of image augmentations to improve the diversity of images that the model sees \citep{10.1093/mnras/stag773}. These include rotations ($0-180^\circ$), zooms (as a centre crop between $0.6-0.7\times424$ pixels)  and vertical flips ($50\%$ rate). Finally, we resize the images to $224\times224$ pixels--this reduces the computational cost of the model and ensures no loss of information in the downsampling steps. We show examples of these images in Figure \ref{fig:rgb_images}.

\begin{figure}
    \centering
    \includegraphics[width=1\columnwidth]{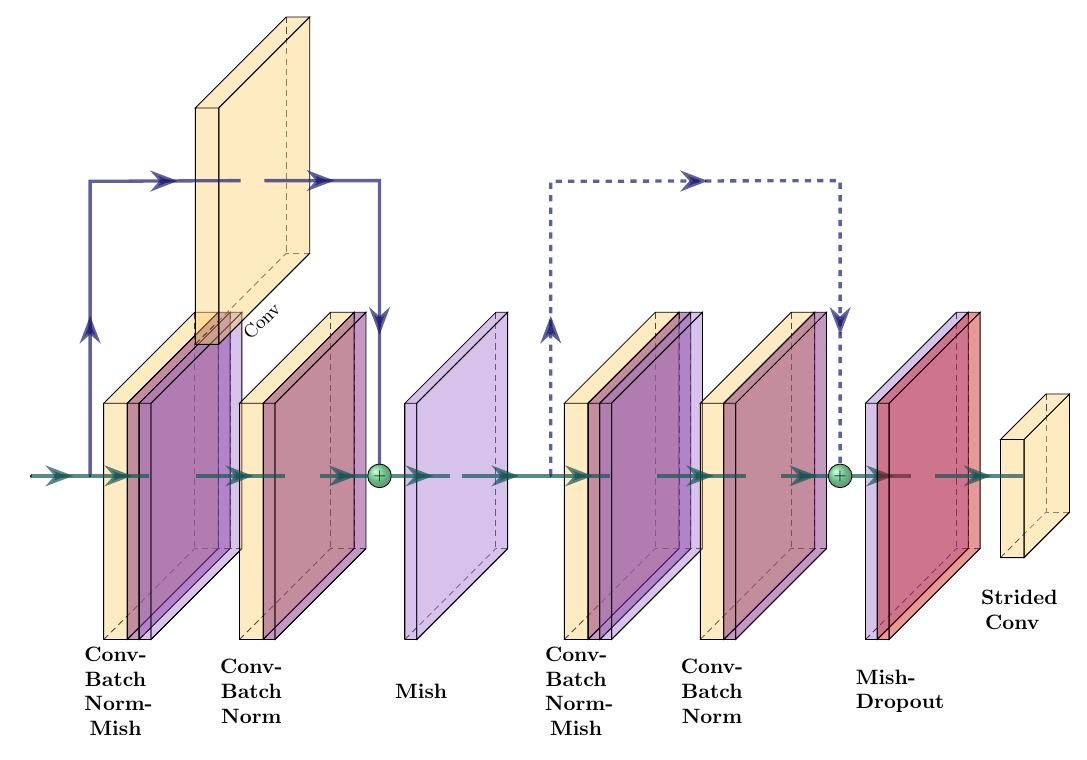}
    \caption{A single step in the encoder arm of ZB:3D. The arrows show the flow of information in a forward pass of the network. There are two skip connections shown, denoted by the arrows splitting off from the forward pass. The solid arrow denotes a residual connection, while the dashed arrow is an identity connection. The plus symbols show where the skip connections are added back into the forward pass. In the decoder, the final strided convolution is replaced by a strided transpose convolution.}
    \label{fig:Arch1}
\end{figure}

\begin{figure*}
    \centering
    \includegraphics[width=1\linewidth]{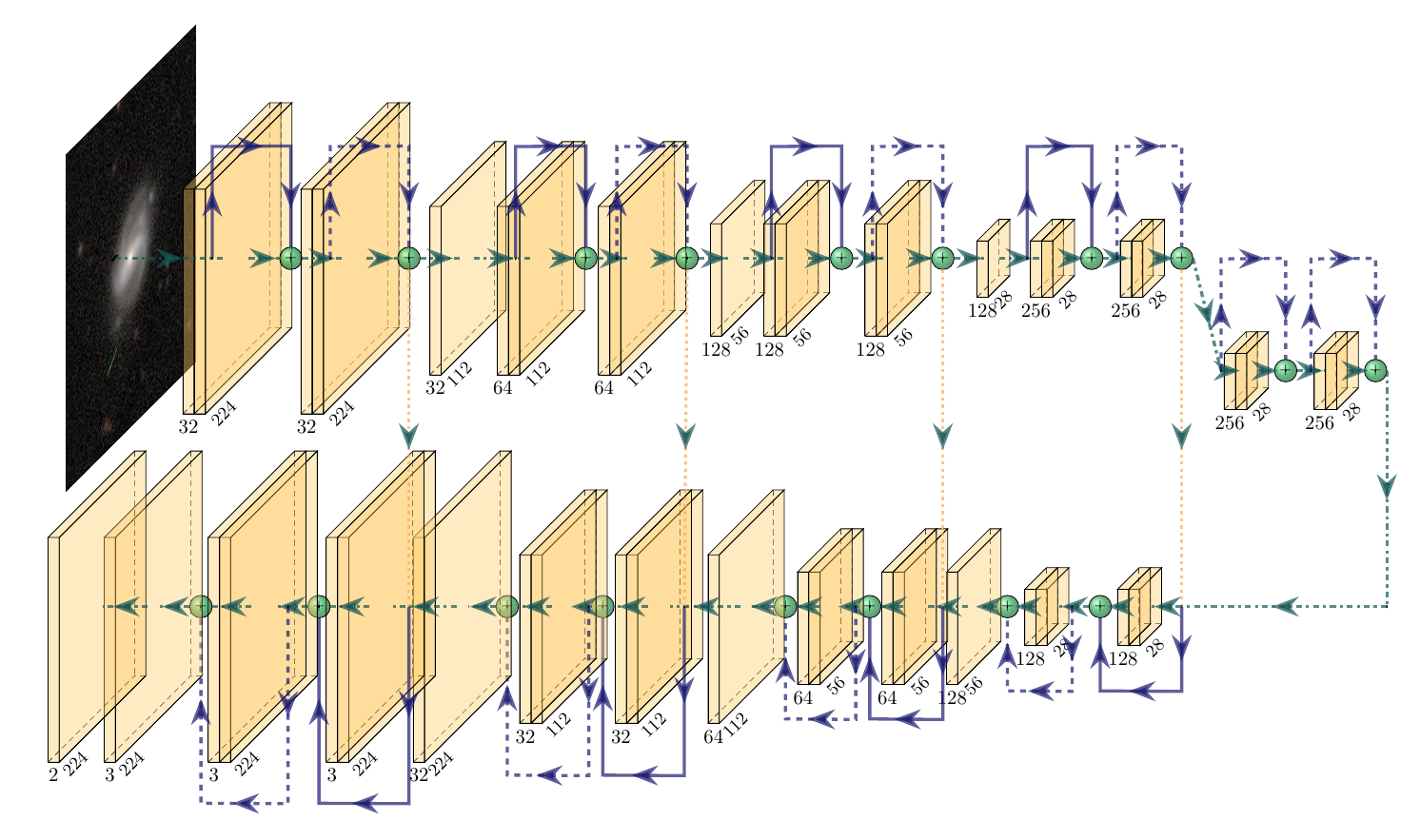}
    \caption{An illustration of the overall architecture of ZB:3D. Each box represents a convolutional layer: the two numbers show the number of filters in each layer, and the number of pixels along one side of the activation maps. The dot-dash arrow shows the main flow of information in a forward pass through the network. The solid arrows represent residual skip connections and the dashed arrows represent identity skip connections. Finally, the dotted arrows show where the encoder step outputs are concatenated with the inputs to each decoder step. (Note that this increases the number of input features to the decoder steps--for example, the input of the first decoder step has $512$ features, $256$ from the previous block of convolutions, and $256$ from the skip connection.) Not shown are the batch normalisation layers, dropout layers, or activation layers.}
    \label{fig:Arch2}
\end{figure*}

\section{ZooBot:3D}
\label{sec:ZooBot}

We use a U-net \citep{ronneberger2015unetconvolutionalnetworksbiomedical} style deep neural network for ZooBot:3D (ZB:3D), which was briefly described in WS23. The core architecture is derived from the U-net in \cite{2022MNRAS.511.1808S}, which was used as part of a denoising diffusion probabilistic model to produce high-quality synthetic galaxy images. U-nets are similar to convolutional autoencoders (see \citealt{Spindler_2020} and references within for a detailed description of these models), but contain skip connections between the outputs of convolutional layers in the downsampling arm of the model to the corresponding layers in the upsampling arm. The two arms share similar structures, relying on ResNet-style residual blocks \citep{he2015deepresiduallearningimage, srivastava2015highwaynetworks}, strided convolutions and transpose convolutions to extract features from the images. The use of residual connections and larger input images allows ZB:3D to be deeper than other works in this area (e.g. \citealt{2023BAAA...64..253R} and \citealt{2024MNRAS.530.1171C}).

A single step in the encoder arm is made up of a residual block, followed by an identity block (see \citealt{he2015deepresiduallearningimage} for definitions of these blocks), followed by a dropout layer to prevent overfitting in all but the first step--the output of the dropout layer is fed into a skip connection to the corresponding step in the decoder. We show a visual representation of this structure in Figure \ref{fig:Arch1}. In the first three steps of the encoder we use a convolutional layer with a stride of $2$ to reduce the size of the receptive field by half. The fourth and final step is fed into two more identity blocks, the last of which serves as the output of the encoder.

The decoder arm follows the same structure, with the notable difference that we use a transpose convolution layer to upsample the activation maps, in place of the strided convolutions in the encoder. It takes three of these steps for the activation maps to reach the original image size. We follow this with a convolutional layer with three output filters to match the channels of the input images. One final convolutional layer produces the output segmentation maps, with a number of filters equal to the desired number of classes--in our case this is two (spiral arms and bars).

We use batch normalisation on all but the final convolution, and the MISH activation function throughout the model \citep{misra2020mishselfregularizednonmonotonic}, except on the final output which is ReLU activated. We illustrate the full architecture of ZB:3D in Figure \ref{fig:Arch2}, which shows the convolutional layers, skip connections, and the dimensions of the activation maps.

The full loss function used in training the model is detailed in WS23. In short, the model predicts, for each pixel, the fraction of volunteers that included that pixel in their mask. We then calculate the mean absolute error between our predicted vote fractions and the real vote fractions, and sum over the full image for each class in the output.

Following a hyperparameter sweep to select the batch size, dropout rates, and down/upsampling dimensions, we train the model with early stopping for a maximum of $5000$ epochs. The final training configurations and model implementation are available at \url{https://github.com/mwalmsley/zoobot-3d}.

During prediction for new segmentation maps, we follow the same data preparation steps for the training data, with the exception that no random augmentations are performed. A fixed zoom factor of $0.65$ is used for prediction images in order to match the cropped dimensions of the training images.

ZB:3D has roughly $3.3$ million parameters, and takes approximately $30$ minutes to predict spiral arm and bar masks for all $639,636$ galaxies in our selected sample when run on a Tesla V100 GPU with 32GB of VRAM.

\begin{figure*}
    \centering
    \includegraphics[width=1.25\linewidth, angle=90]{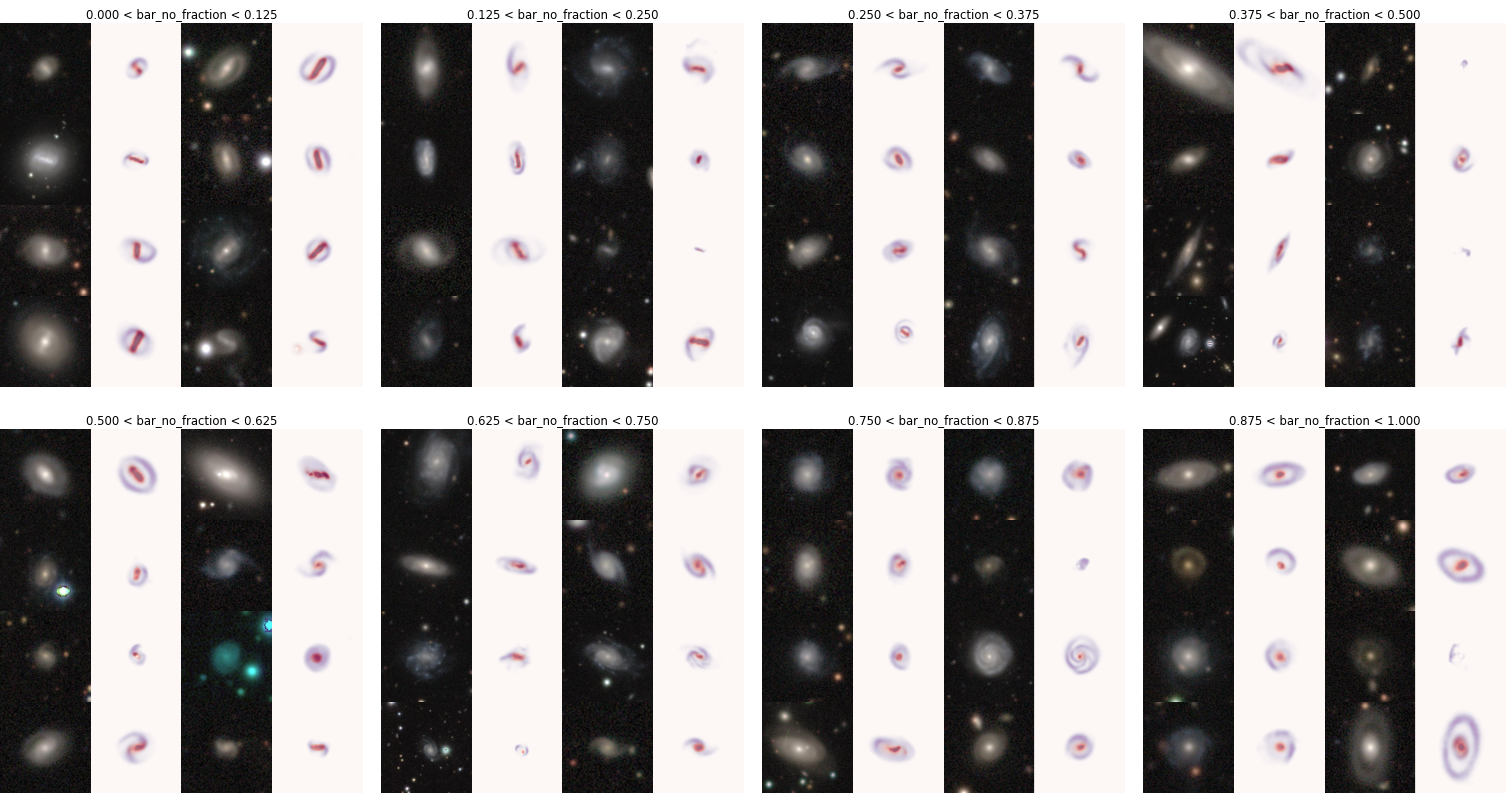}
    \caption{Galaxies and their segmentation maps, predicted by ZB:3D. For each galaxy we show the DESI RGB image, followed by the a visualisation of the segmentation maps, with the spiral arms in purple and the bars in red. Galaxies are divided into groups based on the probability that the galaxy does not contain a bar, as predicted by GZ: DESI. Each subplot title indicates the range of bar probabilities, and the galaxies are chosen at random from that range.}
    \label{fig:ZB3D_seg_maps_bar}
\end{figure*}

\begin{figure*}
    \centering
    \includegraphics[width=0.8\textwidth]{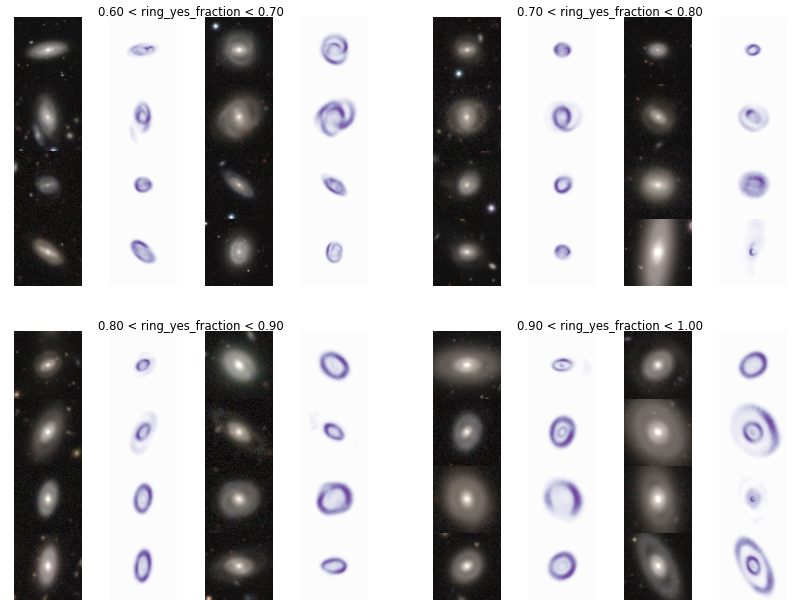}
    \caption{A sample of ring galaxies from DESI, selected using GZ: Rings, processed using the segmentation model. The segmentation maps shown are predicted by the spiral arm part of the model. TK: remake with random selection without doubles.}
    \label{fig:ZB3D_rings}
\end{figure*}


\begin{figure*}
    \centering
    \includegraphics[width=0.8\textwidth]{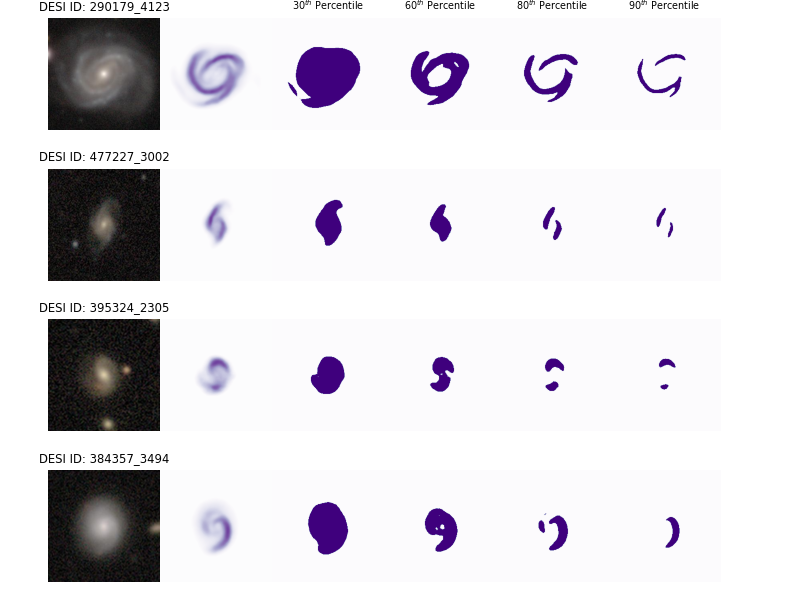}
    \caption{Segmentation thresholds utilising percentile cuts in a given segmentation map for four spiral galaxies. We show the DESI RGB image in the left-most column, followed by the ZB:3D map of the spiral arms and four example pixel thresholds set at the $30^{th}$, $60^{th}$, $80^{th}$ and $90^{th}$ percentiles.}
    \label{fig:ZB3D_percents}
\end{figure*}

\begin{figure*}
    \centering
    \includegraphics[width=0.8\textwidth]{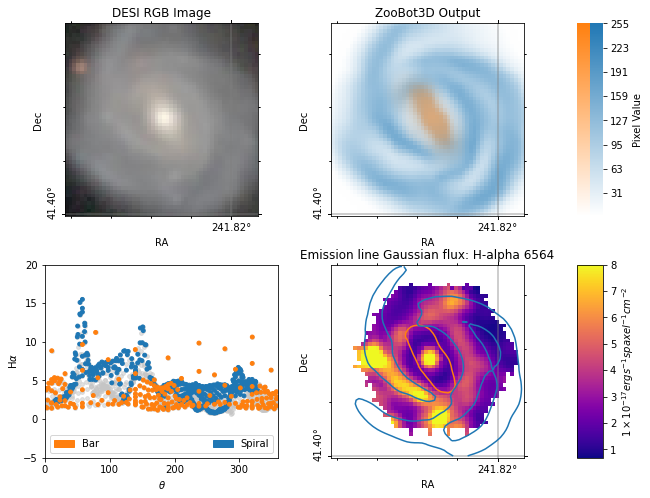}
    \caption{An example of applying the ZB:3D segmentation maps to a MaNGA IFU (MaNGA-ID: $1$-$248420$, also shown in Figure 6 of \citealt{2021MNRAS.507.3923M}). The top left panel shows the DESI RGB image, and the top right shows the ZB:3D spiral arm and bar masks. Both of these plots are projected into the wcs frame of the MaNGA datacube. The bottom left panel shows the azimuthal distribution of $H_{\alpha}$ flux for spaxels in the spiral arms (purple), bar (orange) and the rest of the disk (grey). The bottom right plot shows the $H_{\alpha}$ flux map, with the spiral arm and bar masks overlayed as contours.}
    \label{fig:ZB3D_IFU}
\end{figure*}

\section{Results}
\label{sec:Results}

In WS23, we showed that ZB:3D produces qualitatively convincing segmentation maps, which were preferred by professional astronomers over the original GZ:3D volunteer maps and segmentation maps extracted using \texttt{SpArcFiRe}. Here, we will show that not only do these maps look convincing across a wide range of galaxy properties, but that they can be used to investigate the physical properties of galaxies.

We begin with Figure \ref{fig:ZB3D_GZ3D_comp}, which shows a comparison of the ZB:3D segmentation maps with the corresponding GZ:3D maps, for three barred spiral galaxies. ZB:3D is capable of identifying the complex structure of the spiral arms in all three galaxies, picking up on details that are not immediately obvious in the volunteer maps. The centre panel of Figure \ref{fig:ZB3D_GZ3D_comp} in particular shows some disagreement in the volunteer maps on the extent of the spiral arms--some volunteers draw tightly around the maxima of the arms, while others cover a wider extent. In the ZB:3D map, we can more clearly see the separation and extent of the spiral arms. All three galaxies show good agreement with the bar masks drawn by the GZ:3D volunteers, likely due to this being a much simpler task with less room for disagreement.

Interestingly, we find that the regression approach used in training ZB:3D results in "fuzzy" maps. The fuzziness appears to be inherited from the input images themselves, with the intensity of the spiral arm mask corresponding to the brightness of the feature in the source image. Other attempts at automated segmentation--which treat the task as a binary classification problem--typically produce maps with sharper edges. This comes with advantages and disadvantages. On the one hand, ZB:3D is able to identify smaller, fainter structures that may be missed by a model trained on maps with a binary threshold applied. On the other hand, the fuzziness introduces some difficulties in deciding which pixels to include when applying the maps to images and IFU data cubes. We will discuss this further in Section \ref{sec:SegThresh}.

To demonstrate ZB:3D's capability to identify structures across a wide range of galaxy morphologies, we provide Figure \ref{fig:ZB3D_seg_maps_bar}. In this figure, we have divided galaxies into bins based on their \texttt{bar\_no\_fraction} from GZ:DESI. From the top left corner to the bottom right, galaxies progress from being either strongly or weakly barred, to being unbarred spirals\footnote{The \texttt{bar\_no\_fraction} here refers to the fraction of volunteers who said that the galaxy does not have a bar, and corresponds to $1 - (\texttt{bar\_weak\_fraction} + \texttt{bar\_strong\_fraction})$}. This figure shows that the model is capable of identifying the spiral arms in a wide range of arm multiplicities, winding angles and clumpiness. Where the model appears to struggle is where one spiral arm is not as bright as the others, or is otherwise less apparent in the source image. The model also produces a noticable output in the bar masks for galaxies where $\texttt{bar\_no\_fraction} > 0.9$, i.e. where there is certainly no bar present; in these cases it appears that the model is instead responding to the central bulge of the galaxy. On the surface, this appears to be a logical outcome of the model responding to the colour and brightness of pixels in the central regions of the image. For this reason, we encourage users of these maps to carefully consider the selection cuts on the samples used in their research.

Most surprising from Figure \ref{fig:ZB3D_seg_maps_bar} are the small number of ring galaxies present in the sample, particularly in the bottom right bin with the highest \texttt{bar\_no\_fraction}. To investigate the model's response to ring galaxies further, we use the classifications from GZ: Rings \citep{Walmsley_2022}. This is shown in Figure \ref{fig:ZB3D_rings}. It should be noted that identifying the extent or shape of rings in galaxies was not part of the original GZ:3D workflow, and so we should not expect a significant training signal for these types of galaxies. Upon cross-matching the galaxies that received volunteer spiral arm masks with the GZ: Rings catalogue, we find only $373$ of the $6,382$ spiral galaxies in our training sample with $\texttt{ring\_yes\_fraction} > 0.7$, and only $3$ galaxies with $\texttt{ring\_yes\_fraction} > 0.85$. Most of these galaxies appear to be pseudo-rings, lenses or tightly wound spirals, as opposed to the true rings seen in Figure \ref{fig:ZB3D_rings} \citep{buta_1996, buta_2017}. Despite this weak training signal, when presented with ring galaxies, ZB:3D produces visually consistent segmentation maps of the galaxy rings. This raises the question of whether this could be considered a form of emergent behaviour, or whether the tasks of labelling spiral arms and rings are just similar enough that it does not require fine-tuning or transfer learning.

\subsection{Segmentation Thresholds}
\label{sec:SegThresh}

In order to utilise the ZB:3D segmentation maps, threshold cuts must be applied to the "fuzzy" outputs of the model. For simplicity, we will refer to the value of any given pixel in the segmentation maps as the "intensity" of that pixel. For more traditional approaches to segmentation, the intensity of a pixel can be considered as a measure of certainty or probability that the pixel is in a given class. However, this is not the case in our regression-based approach, where we are predicting the confidence of the crowd, as opposed to a binary class label. Each galaxy can have a wide range of pixel intensities, with some galaxies having pixels with much higher intensities than others. This can be seen in Figure \ref{fig:ZB3D_GZ3D_comp}, where the spiral arm masks of the top and bottom panels are somewhat darker than the middle panel.

This variance in the pixel intensities means that choosing an appropriate threshold value for the spiral arm masks is non-trivial. In GZ:3D, it was suggested that a pixel could be included in the spiral arm map if three or more volunteers included the mask. However, we find that using the same fixed value for the threshold produces very inconsistent results, due to the varying ranges of intensities between different galaxies.

Instead, we investigated methods of selecting a threshold value, based on the properties of the intensity distribution for a given galaxy. We considered both cuts based on fractions of the maximum intensity of pixels in the given map, and the percentiles of the intensity distribution. We found that both methods work well, with the percentile cuts acting slightly more consistently when cutting out the galaxy centres and inter-arm regions. Figure \ref{fig:ZB3D_percents} shows a range of possible percentile cuts, from a fairly generous $30^{th}$ percentile cut which encompasses most of the disk, to stricter cuts at the $80^{th}$ and $90^{th}$ percentiles.  Throughout the rest of this paper, where spiral arm masks are applied, we will apply a cut at the $80^{th}$ percentile of the intensity distribution, unless stated otherwise.

We find less variation in the intensities of pixels in the bar masks produced by ZB:3D. However, to remain consistent, we also utilise a cut at the $80^{th}$ percentile in the intensity distribution for the bars as well.

\subsection{Application to IFU Data Cubes}

When combined with integral field spectroscopy, segmentation mapping can allow for detailed comparative analysis of galaxy structures, illuminating the roles they play in galaxy evolution.

As a first test of the ZB:3D maps, we repeat the analysis done in GZ:3D for MaNGA galaxy MaNGA-ID $1-248420$. In Figure \ref{fig:ZB3D_IFU} we show the azimuthal distribution of $H_{\alpha}$ emission, from the MaNGA Data Analysis Pipeline MAPS file \citep{Belfiore_2019, Westfall_2019}. We highlight the extent of the bar and spiral arms with contour lines, which shows that the spiral arm masks line up very well with the peaks in $H_{\alpha}$ emission in the disk. The bar mask also lines up with the so-called "star formation desert" region around the centre of the galaxy, where the bar has swept out a region devoid of cold gas, which has been funnelled into the galaxy centre.

\subsection{Application to Photometric Data}

Applying the segmentation maps to photometric imaging allows for significantly larger datasets of galaxies to be analysed, though with limitations compared to IFU data in the complexity of the data available. We demonstrate two simple use cases for the data products by comparing the broadband colours of pixels within the galaxy structures and in the rest of the disk.

\begin{figure}
    \centering
    \includegraphics[width=1\linewidth]{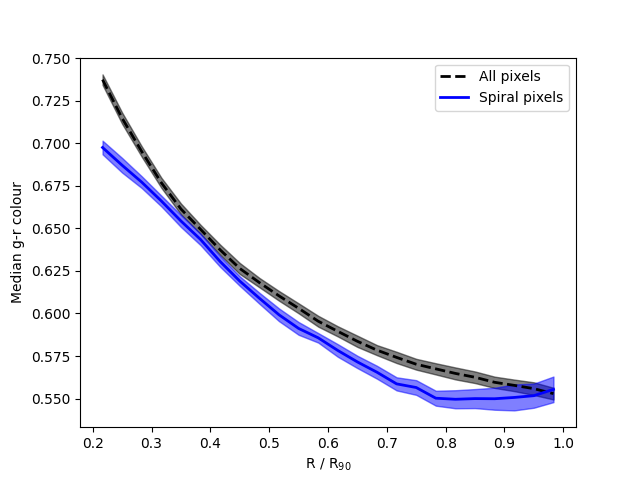}
    \caption{The median $g-r$ colour in bins of effective radii for $16,413$ galaxies with \texttt{has\_spiral\_arms}$>0.985$. For each galaxy, we calculate the mean $g-r$ colour of all pixels in each radial bin (dashed black line), as well as the mean $g-r$ colour of just the pixels labelled as being part of the spiral arms (solid blue line). Each radial profile represents the median from $1000$ bootstrap resamples, and the shaded regions show two standard deviations in the bootstrap median.}
    \label{fig:ZB3D_radial_colour_all}
\end{figure}

Spiral arms are known to concentrate star formation within disk galaxies, though the exact mechanism is still debated, resulting in the spiral arms appearing bluer than the rest of the disk in optical imaging \citep{1969ApJ...158..123R, 2021ApJ...917...88Y, Sellwood_2022}. In Figure \ref{fig:ZB3D_radial_colour_all}, we show how the colours of a galaxy disk vary between the peaks of the spiral arms and the disk overall, for galaxies with a high certainty of containing spiral arms in the GZ-DESI classifications (with $\texttt{has\_spiral\_arms} > 0.985$). We calculate the colours using the source $grz$ FITS images from DESI. For this purpose, we reproject the segmentation maps into the original WCS frame of the DESI cutouts. To calculate the individual radial profiles of each galaxy, we take $25$ radial bins between $0.2 R_{90}-1R_{90}$ to avoid the inner bulge region, and find the mean $g-r$ colour of all pixels (including spiral arm pixels) in the bin and of just those pixels contained in spiral arm masks. We then use bootstrap resampling to find the statistical mean and standard deviation of the median profile of all galaxies in the subsample, across 1000 samples with replacement. As expected, we find that the median colour for pixels within the spiral arm masks is notably bluer than the rest of the pixels within the galaxy disks at all but the most extreme radii (out to $1 R_{90}$). This result agrees with the example science case shown in Figure $7$ of \cite{2021MNRAS.507.3923M}, which shows enhanced star formation in spiral arms using MaNGA data.

\begin{figure}
    \centering
    \includegraphics[width=1\linewidth]{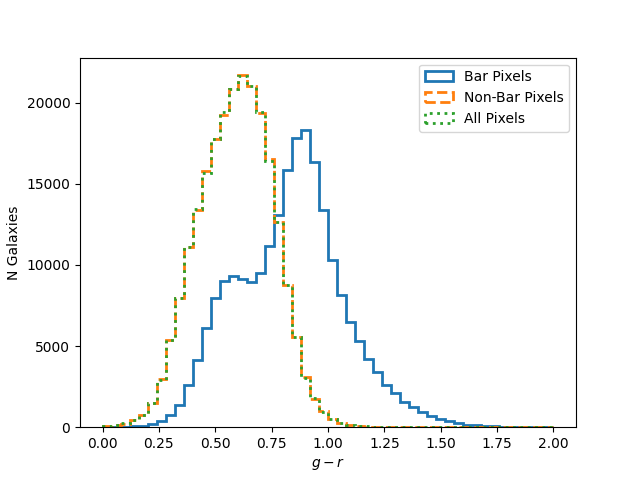}
    \caption{Comparison of the $g-r$ colours of the bar regions identified by the segmentation model, and the rest of the disk, for $230,511$ barred galaxies. We show the median $g-r$ colours of pixels in the bar with a solid line, the median of the non-bar pixels with the dashed line, and the median from the full disk with the dotted line.}
    \label{fig:ZB3D_bar_colour_comp}
\end{figure}

Similar to spiral arms, we would also expect the bars of barred galaxies to affect the colour of pixels in the cutouts. Bars are typically redder than the rest of the disk \citep{2018MNRAS.473.4731K}, and we show this to be the case in Figure \ref{fig:ZB3D_bar_colour_comp}. As with the previous figure, we reproject the bar masks of the $230,511$ barred galaxies in the dataset into the original DESI WCS frame of reference. We then calculated the median pixel $g-r$ colour for bar pixels, non-bar pixels and all pixels in the disk. This figure shows a very clear excess toward redder colours for pixels within the bar masks. Comparing Figures \ref{fig:ZB3D_radial_colour_all} and \ref{fig:ZB3D_bar_colour_comp}, we can also see that the excess in the barred pixels is significantly redder than the mean $g-r$ colours at all radii in spiral galaxies.

\subsection{Data Model}
\label{datamodel}

We have made the dataset publicly available via Zenodo. To aid in usability of these data products, we have chosen to keep the file structure relatively simple. For each galaxy processed by ZB:3D there is one \texttt{.fits} file, which contains:

\begin{itemize}
\item {\tt HDU 0: [image]} DESI-LS cutout as seen by ZB:3D at time of prediction. The images are prepared as described in Section \ref{sec:Images}, and we use a fixed zoom factor of $0.65$ as noted in Section \ref{sec:ZooBot}, resulting in a $224 \times 224$ pixel RGB image.
\item {\tt HDU 1: [image]} The predicted spiral arm mask from ZB:3D, with pixel intensities stored as $8$-bit unsigned integers.
\item {\tt HDU 2: [image]} The predicted bar mask from ZB:3D, with pixel intensities stored as $8$-bit unsigned integers.
\end{itemize}

We include the wcs coordinates for the transformed images in the headers of the \texttt{.fits} file, to allow the masks to be reprojected into relevant frames of references, for example into the wcs frame of the MaNGA data cubes.

\section{Conclusions}
\label{sec:Conclusions}

We have presented a new dataset of machine-generated segmentation maps for disk-type galaxies in the DESI-LS. We trained a deep U-Net using the volunteer-labelled spiral arm and bar pixel masks from Galaxy Zoo: 3D, and generated corresponding segmentation maps for 639,636 galaxies. This sample includes 512,216 galaxies with a high likelihood of having spiral arms, and 230,511 galaxies with a high likelihood of having a strong or weak bar, according to the classifications from GZ-DESI. We also include $29,006$ objects from the MaNGA survey and $5,019$ objects from the SAMI survey.

The segmentation maps provide an accurate "soft" representation of the spiral arms and bars, achieved by using the GZ:3D vote fractions at the regression target, as opposed to binary segmentation labels. The maps show good agreement with the GZ:3D volunteer labelled maps, and qualitatively agree with the observed structures inside the sample galaxies. We have used the imaging from DESI-LS as the input training images, as opposed to the SDSS images used in GZ:3D, as the deeper imaging provides more detail into the internal galaxy structures. This decision also allows us to apply the model to a much wider range of objects.

Despite the fact that identifying galactic rings was not part of the original GZ:3D task, we observe that the spiral arm section of the model's segmentation head does trace out the extent of these structures. Using the classifications from GZ:Rings, we have shown that the model is capable of identifying rings within the main disk of the target galaxies and nuclear rings in the bulge region. This behaviour opens the door to studying the differences in properties between the rings and the inner/outer ring regions of galaxies. Further improvements could be made to these classifications by running a small citizen science project to label rings as a new training set.

As the segmentation maps generated by ZooBot:3D represent the confidence of the model that a given pixel lies within the spiral arms or bars, it is necessary to turn the maps into binary masks so that they can be used to isolate the individual structures. To do this, we investigate two schemes of applying cuts to the pixel values to create these binary masks: a cut on the maximum pixel value of the map, and a percentile cut based on pixels within the galaxy disk. We find that both schemes produce good results, and opt for a percentile cut at the $80^{th}$ percentile for our own analysis--we encourage users of the data products to experiment with their own cuts to fit their research goals.

Finally, we have demonstrated how the data products can be used with both IFU spectroscopy and photometric imaging. Following the analysis performed in \cite{2021MNRAS.507.3923M} on the GZ:3D pixel masks, we have shown that our new maps can be used to separate the spiral arm and bar regions in MaNGA datacubes, enabling measurements of emission lines, spectral properties and derived properties of galaxies observed with IFU spectroscopy. Using the DESI-LS photometric imaging, we have reproduced the known relationships in colours between spiral arms, bars and the galaxy disk in total. Radially, we find that the spiral arm regions are bluer than the disk overall, which agrees with findings that star formation is enhanced in the inner-arm regions. In barred spirals, we find that the bar regions are, on average, redder than the disk overall, which also agrees with previous studies.

A century on from the first classifications of galaxy morphology, many mysteries remain in our understanding of the formation and role of these iconic structures. The availability of these new data products will aid in observational tests and statistical studies of disk galaxies and help answer fundamental questions about their origins. In this new era of astronomy, with imaging from Rubin, Euclid and the forthcoming Roman telescope, segmentation can play an increasingly important role by facilitating decomposed and structural studies of potentially millions of galaxies.

\section*{Acknowledgements}

This research has made use of the University of Hertfordshire high-performance computing facility.

The Dunlap Institute is funded through an endowment established by the David Dunlap family and the University of Toronto.

ILG has received the support from the Czech Science Foundation Junior Star grant no. GM24-10599M.

BDS acknowledges support through a UK Research and Innovation Future Leaders Fellowship [grant number MR/T044136/1] and its renewal [grant number MR/Z000076/1].

JJP acknowledges funding from the Science and Technology Facilities Council (STFC) Grant Code ST/X508640/1.

\section*{Data Availability}

The code for ZB:3D is publicly available via GitHub at \url{https://github.com/mwalmsley/zoobot-3d}.

We also provide a public respository of the outputs of this data, and the sample catalogues associated with them, at Zenodo. This includes the sample selection for the 639,636 galaxies from GZ:DESI, 29,006 galaxies from MaNGA and 5,019 galaxies from SAMI. The repository also contains the \texttt{.FITS} files described in Section \ref{datamodel}. The \texttt{.FITS} files for the main data release are contained in the \texttt{.zip} files with filenames in the format \texttt{ZooBot3D-??.zip}. The MaNGA and SAMI samples are found in the files with names \texttt{MaNGA-??.zip} and \texttt{SAMI-??.zip}, respectively.



\bibliographystyle{mnras}
\bibliography{example} 



\bsp	
\label{lastpage}
\end{document}